\documentclass[nojss]{jss}
\usepackage{amsmath}
\usepackage{arydshln}

\author{David Randahl \\ Dep. of War Studies \\ Swedish Defense University \And Anders Hjort \\ Eiendomsverdi AS \And Jonathan P. Williams \\ Dep. of Statistics \\ North Carolina State University}
\title{\pkg{pintervals}: an \proglang{R} package for model-agnostic prediction intervals}

\Plainauthor{David Randahl, Anders Hjort, Jonathan P. Williams} 
\Plaintitle{pintervals: an R package for model-agnostic prediction intervals} 
\Shorttitle{\pkg{pintervals}} 

\Abstract{
The \pkg{pintervals} package aims to provide a unified framework for constructing prediction intervals and calibrating predictions in a model-agnostic setting using set-aside calibration data. It comprises routines to construct conformal as well as parametric and bootstrapped prediction intervals from any model that outputs point predictions. Several R packages and functions already exist for constructing prediction intervals, but they often focus on specific modeling frameworks or types of predictions, or require manual customization for different models or applications. By providing a consistent interface for a variety of prediction interval construction approaches (all model-agnostic), \pkg{pintervals} allows researchers to apply and compare them across different modeling frameworks and applications.
}
\Plainkeywords{prediction intervals, conformal prediction, bootstrapping, R,} 

\Address{
  David Randahl \\ Dep. of War Studies \\
  Swedish Defense University \\
  Drottning Kristinas Väg 47 \\ 114 28 Stockholm, Sweden \\
  E-mail: \email{david.randahl@fhs.se}\\
  URL: \url{http://randahl.org/david}
}

\begin{document}


\section{Introduction}

Forecasting has in recent years become an increasingly central component of quantitative research in the social sciences. From predicting civil conflict and mass atrocities to anticipating electoral outcomes, economic downturns, and policy adoption, social scientists are increasingly engaged in producing predictive models for a range of empirical phenomena \citep[see for instance][]{hegre2017introduction, nanlohy2017policy, lyu2021forecasting, hegre20242023, petropoulos2022forecasting}. This shift reflects a growing recognition that forecasting can complement traditional causal inference by providing actionable, forward-looking insights to policymakers, civil society, and researchers alike \citep{nanlohy2017policy}.

Most of these predictive models provide a single estimate of an outcome of interest, such as the expected number of conflict-related fatalities in a country-month \citep[e.g.~][]{hegre2019views, hegre2021views2020,hegre20242023}, or the probability of electoral violence in a certain election \citep[e g.][]{randahl2025forecasting}. While such point predictions can offer valuable insights they do not reflect the inherent uncertainty of the predictions, limiting utility and potentially leading to overconfidence in the forecasts. In many applications, understanding the uncertainty around point predictions is crucial for effective decision-making, as it provides a more complete picture of the risks and potential outcomes associated with different scenarios \citep{raftery2016use}. For instance, in conflict forecasting, it may be more important for stakeholders, in order to effectively prioritize their resources, to know the likelihood that a conflict exceeds a certain threshold of fatalities, rather than just the expected number of fatalities. Similarly, the width of a prediction interval informs decision-makers about the level of confidence they may place in a particular point prediction, which is essential for risk assessment and planning.

These limitations have led to a growing consensus that point predictions alone are not sufficient for effective communication with stakeholders or for providing actionable support to decision-makers, especially in high-stakes domains such as conflict forecasting. When forecasts inform real-world interventions, it is not enough to know what the most likely outcome is; decision-makers must also understand how much uncertainty encompasses the most likely outcome, and how much confidence they can place in it \citep{raftery2016use,hegre20242023, brandt2014evaluating, chadefaux2017conflict}.  For example, a forecast suggesting a moderate risk of violence may warrant very different policy responses depending on whether the associated uncertainty is low or high. Transparent communication of predictive uncertainty can therefore help ensure that responses are proportionate to the level of risk, improving both reliability and accountability.

To quantify the uncertainty around predictions, researchers commonly turn to prediction intervals. A variety of techniques exist for constructing such intervals, including parametric methods (i.e.~assuming a specific distribution for the prediction errors), bootstrapping (resampling errors from a calibration set), quantile regression, and Bayesian credible intervals. Each approach has strengths and limitations. Parametric intervals, when available analytically or approximately, may be easy to compute but rely on strong distributional or asymptotic assumptions. Bootstrap methods are flexible, but they assume that the distribution of errors is constant across the data, and they tend to underestimate variance (due to sampling from an empirical distribution function rather than the true cumulative distribution function). Quantile regression provides asymmetric intervals but requires careful specification and is model-specific. Bayesian methods allow for full posterior distributions but are also model-specific and require careful prior specification  \citep[for a review of these methods, see][]{tian2022methods, zhang2020random, hesterberg2015teachers}. Second, Bayesian uncertainty is epistemic rather than aleatory, meaning that credible sets cannot necessarily be interpreted as [frequentist] confidence sets; and third, Bayesian posterior inference is subject to the {\em false confidence theorem} \citep{balch2019,martin2019,carmichael2018}, meaning that there always exists sets not containing the true value-to-be-predicted, that are assigned arbitrarily large posterior probability, arbitrarily often (i.e.~over repeated sampling of data sets). The limitations of these standard methods may result in unreliable/misleading uncertainty quantification, particularly in complex or heterogeneous data settings common in social science applications.  There has thus been a demand for more-flexible, model-agnostic approaches to uncertainty quantification that offer verifiable reliability guarantees, and that can be applied across a wide range of predictive models and data types.  To meet this demand, conformal prediction (CP) is proving to be a viable approach to uncertainty quantification.

Under the assumption of exchangeable data observations\footnote{Exchangeability, requiring only that the joint distribution of the data be invariant to permutations, is a slightly weaker assumption than the data being independent and identically distributed.}, CP produces prediction sets with guaranteed coverage for any fixed sample size, at any chosen confidence level. A key advantage of CP is its model-agnostic nature: it can be applied on top of any prediction model, from simple linear regression models to complex machine learning algorithms, without requiring model-specific modifications \citep{shafer2008tutorial, vovk2022algorithmic}. Recent extensions, including clustered CP \citep{ding2023class, hjort2024clustered}, weighted CP \citep{hjort2025uncertainty}, and bin-conditional CP \citep{randahl2025bin}, have further enhanced its applicability by addressing local calibration challenges in settings with skewed or imbalanced outcome distributions.  

This article serves as an introduction to the \pkg{pintervals} package, and as an expository on the theoretical and computational foundations of the supported methods for uncertainty quantification. We also showcase the package using data on county election turnout in the United States presidential election of 2016. The package is designed to be flexible and extensible, allowing users to apply it to a wide range of predictive modeling tasks in the social sciences and beyond.

Several R packages and functions already exist for constructing prediction intervals. For example, the \pkg{forecast} package \citep{forecastpkg} provides methods for time series forecasting, while the \pkg{tidymodels} and \pkg{mlr3} packages \citep{tidymodelspkg, mlr3} offer tools for machine learning workflows. Parametric prediction intervals can be constructed using the \pkg{stats} package \citep[for linear and generalized linear regression models;][]{statspkg}, or by manually computing intervals (when analytically available) based on standard distribution functions (e.g.~\code{qt} for t-distribution, \code{qnorm} for normal distribution, etc.). Bootstrapped prediction intervals can be constructed using the \pkg{boot} package \citep[namely, it provides functions for resampling;][]{bootpkg}, or by manually implementing bootstrapping procedures using the \pkg{stats} package. Conformal prediction intervals can also be obtained through a range of packages, such as the \pkg{conformalbayes} package \citep{conformalbayespkg} for Bayesian conformal prediction, the \pkg{conformalForecast} package \citep{conformalforecastpkg} for time series forecasting, and the \pkg{LCP} package \citep{lcppkg} for localized conformal prediction. Beyond these packages, there are also various tutorials on how to obtain conformal prediction intervals manually using functions available in packages such as \pkg{tidymodels} \citep{tidymodelspkg}, and \pkg{marginaleffects} \citep{marginaleffectspkg, arel2025model}.

While existing packages offer valuable tools for constructing prediction intervals, they generally lack a unified, model-agnostic interface. Most are tied to specific modeling frameworks or require extensive manual adjustments to accommodate different models or applications, making it challenging for researchers to apply and compare them consistently across datasets and prediction tasks. The \pkg{pintervals} package addresses this limitation by providing a coherent and model-agnostic framework for generating prediction intervals. It implements several widely used modern and classical approaches, including CP (with subcategories such as weighted, clustered, and bin-conditional CP), bootstrap-based intervals, and parametric methods, and is compatible with any model that produces point predictions. \pkg{pintervals} utilizes a structure where separate calibration data are used to estimate the uncertainty around predictions, allowing for flexible and robust uncertainty quantification.

\section{Prediction intervals}

Prediction intervals quantify the uncertainty around model predictions by providing a range within which the true outcome is stated to be contained at a given confidence level, e.g.~a $90\%$ prediction interval. Unlike confidence intervals, which characterize uncertainty about fixed but unknown model parameters, prediction intervals capture the uncertainty in subsequent data realizations, making them an essential tool for evaluating and communicating the reliability of forecasting models. In applied research, particularly in the social sciences, this distinction is crucial, as stakeholders and policymakers are often more interested in the uncertainty of predicted outcomes than in parameter inference.      

The \pkg{pintervals} package implements three classes of model-agnostic prediction intervals: conformal, bootstrapped, and parametric. These methods differ in their underlying assumptions and estimation strategies but share a common interface and workflow. Both the conformal and bootstrap approaches rely on a separate calibration set containing observed outcome values, used to calibrate prediction intervals based on the empirical distribution of prediction errors. Parametric intervals, in contrast, assume a specified distribution for the prediction errors of which the parameters can either be estimated from calibration data or supplied directly. Together, the software package provides a flexible and coherent framework for quantifying predictive uncertainty across a wide range of models and applications.

\subsection{Conformal prediction}

CP is a general, distribution-free method that produces prediction intervals with guaranteed marginal coverage, with minimal assumptions on the underlying prediction model. The method relies on the construction of a \textit{non-conformity score} to quantify a degree of dissimilarity of a given datum point versus a collection of data examples. Under the assumption that non-conformity scores are exchangeable, their empirical distribution can be used to construct a p-value for testing the null hypothesis that the non-conformity score of a test point is exchangeable with the observed data scores, e.g.~implying whether or not to rule out the given test point as a plausible predicted outcome.  The conformal p-value is a usual frequentist p-value, in that it is stochastically no smaller than a uniform(0,1) random variable, it gives finite-sample control of type 1 errors, and it corresponds to a prediction interval.  Namely, a CP interval at level $1-\alpha$ contains all outcome values for which their conformal p-value exceeds $\alpha$.  For a comprehensive background on CP, see \cite{vovk2022algorithmic}, and for its connections to other statistical paradigms, see \cite{williams2025,williams2024decision}.

The CP algorithm constructs a prediction interval for $y_{N+1}$ given a corresponding feature $x_{N+1}$, by comparing to a training set of feature-response pairs $(x_1,y_1), \dots, (x_N, y_N)$. The non-conformity score $s_i$ may measure the dissimilarity between a feature-response pair $(x_i, y_i)$ and the remaining data points $(x_1, y_1), \dots, (x_{i-1}, y_{i-1}), (x_{i+1}, y_{i+1}), ..., (x_{N+1}, y_{N+1})$. Although the distribution of the data and of the non-conformity scores are both unknown, assuming exchangeability implies that $s_{N+1}$ is equally likely to take any rank among the $N+1$ scores, i.e.~$\text{rank}(s_{N+1}) \sim \text{uniform}\{1,\dots,N+1\}$. Consequently, if $\hat{q}_{1-\alpha}$ denotes the $1-\alpha$ empirical quantile of the calibration scores, then, assuming exchangeability, $\mathsf{P}(s_{N+1} \leq \hat{q}_{1-\alpha}) \geq 1-\alpha$. In other words, $\text{rank}(s_{N+1})/(N+1)$ is a \textit{p-value} of the null hypothesis that $s_{N+1}$ is indeed exchangeable with the $s_1, \dots, s_{N}$. A CP set, at any level $1-\alpha \in [0,1]$, can be assembled by considering all values that $y_{N+1}$ could take and only including those for which $\text{rank}(s_{N+1})/(N+1) > \alpha$.

In a supervised setting, we typically create a prediction model $\hat{f}$ such that $\hat{f}(x_{N+1})$ represents the model's best guess of $y_{N+1}$. In a regression setting, the canonical non-conformity score is perhaps the absolute residual, $s_i := \lvert y_i - \hat{f}(x_i)\lvert$, while in classification a standard choice is {\em one minus the softmax probability} assigned to the true class. In either case, the CP procedure yields a marginal prediction set
\begin{align*}
    C_{1-\alpha}(x_{N+1}) := \big\{ y \in \mathcal{Y} \, : \, s_{N+1}(y) \leq \hat{q}_{1-\alpha}\big\}, 
\end{align*}
where $s_{N+1}(y)$ denotes the non-conformity score corresponding to $y_{N+1} = y$.  This construction ensures the following finite-sample marginal coverage guarantee, 
\begin{align*}
    \mathsf{P}\{Y_{N+1} \in C_{1-\alpha}(X_{N+1})\} \geq 1-\alpha,
\end{align*}
which holds with virtually no assumptions on the underlying prediction model or choice of non-conformity score. 

There are two main variants of the CP algorithm: the transductive (or \textit{full CP}) and inductive (or \textit{split CP}) algorithm. The transductive utilizes the entire training data, making it maximally statistically efficient, but (typically) requires re-training of the prediction model for each test value $y$, making it computationally costly. The inductive version trades off some share of statistical efficiency for a gain in computational efficiency, by setting aside a portion of the data as a calibration set, untouched during training, but requiring only a single training of the prediction model.  The resulting prediction intervals constructed are typically evaluated based on empirical coverage on a test set, with reference to the nominal $1-\alpha$ confidence level, for each $\alpha \in [0,1]$; if the {\em coverage gap} exceeds usual Monte Carlo simulation variation, then the implication is that the non-conformity scores are {\em not} exchangeable.  In the latter case, alternative formulations of the non-conformity measure and/or prediction model may better account for structural components of the data, to yield exchangeable scores.  In practical applications, the size of the prediction intervals are also of interest, as ``tighter'' intervals are more informative to the user.

\subsubsection{Mondrian conformal prediction}

While standard CP provides valid coverage guarantees marginally, it does not account for potential heterogeneity in the data, which can lead to over- or under-coverage in certain regions of the data. To address this, Mondrian CP (MCP), or label-conditional CP, extends the conformal framework by partitioning the calibration set into groups based on covariates or other characteristics. The disjoint groups enable the construction of group-conditional prediction intervals, i.e.~resulting in a CP set for each group. Supposing the feature space $\mathcal{X}$ can be partitioned into disjoint subsets $\mathcal{X}_1, \dots, \mathcal{X}_K$, MCP guarantees, for each $k \in \{1, \dots, K\}$ and $\alpha \in [0,1]$, that 
\begin{align*}
    \mathsf{P}\{Y_{N+1} \in C_{1-\alpha}(x_{N+1}) \mid x_{N+1} \in \mathcal{X}_k\} \geq 1-\alpha.
\end{align*}
In MCP, the calibration step is performed separately per group rather than across the entire calibration data, involving the computation (for each $\alpha$) of $K$, $1-\alpha$ level quantiles $\hat{q}_{1-\alpha}^{(1)}, \dots, \hat{q}_{1-\alpha}^{(K)}$.  While group-conditional validity may hold in MCP, the limiting factor for efficiency (e.g.~narrow intervals) is that a sufficient number of calibration points must be available in each group. Finally, it is known \citep{LeiWasserman2013} that exact conditional coverage guarantees, i.e.~achieving valid coverage conditional on the exact value $x_{N+1}$ rather than on a Mondrian group membership, are impossible in the context of CP.

\subsubsection{Clustered conformal prediction}

Clustered conformal prediction (CCP) is an extension of MCP for situations where the number of groups $K$ is large and/or some groups contain too few calibration points to efficiently estimate the required quantile. The approach, first proposed in \cite{ding2023class}, is to cluster the $K$ initial groups into $M << K$ clusters, and perform calibration at the cluster level rather than at the group level. By pooling information across similar groups, CCP increases the statistical efficiency and may also help to avoid numerical issues arising from too few examples per group.

The key idea of CCP is to cluster together groups whose empirical distribution of non-conformity scores are similar. As shown in \cite{ding2023class}, the expected coverage gap for any given group depends directly on the similarity between the groups clustered together (note that CCP guarantees validity at the cluster level). When the empirical cumulative distribution functions clustered together are similar, the coverage gap is small. In practice, clustering can be done, for example, through k-means-clustering or based on the Kolmogorov-Smirnov distance between empirical cumulative distribution functions \citep{ding2023class, hjort2024clustered}.

\subsubsection{Distance weighted conformal prediction}
Another modification of the CP procedure is a class of methods referred to as \textit{weighted} CP. Rather than calculating the quantile of interest once (or once per group or cluster), weighted CP calculates $\hat{q}_{1-\alpha}$ as a \textit{weighted} quantile, where the scores $s_{1}, \dots, s_{N}$ are weighted according to some weights $w_{1}, \dots, w_{N}$. By calculating a weighted quantile, the method places more emphasis on non-conformity scores that are determined closer to exchangeable with the test instance, e.g.~with reference to time or space. For example, a weighted CP framework could take weights of the form $w_i \propto \exp\{-\text{dist}(x_i, x_{N+1})\}$, where $\text{dist}(x_i, x_{N+1})$ is an application-specific distance function quantifying the distance between calibration and test features. The prediction interval for $y_{N+1}$ is then calibrated using the weighted $1-\alpha$ quantile of the scores. 

The literature contains several variations of this, including data-driven weights to account for covariate shift \citep{Tibshirani2019}, fixed weights to handle non-exchangeable data \citep{Barber2023}, spatial weights \citep{MartinMaoReich2023}, and feature-distance weights \citep{guan2023localized}. The coverage guarantees of the method depend on the choice of weighting function; see \cite{barber2025unifying} for a review of the statistical properties of weighted CP methods.

\subsubsection{Bin-conditional conformal prediction}

It is known that the standard CP methods often under-cover in lower density regions of the outcome space $\mathcal{Y}$, while over-covering in higher density regions of $\mathcal{Y}$ \citep{guan2023localized}. While MCP and CCP are useful for addressing under- and over-coverage for different groups in the data, these methods are not easily applicable if the outcome space has no natural or well-defined grouping structure. One example is data arising from skewed distributions. Bin-conditional conformal prediction (BCCP) addresses this issue by partitioning the calibration data into user-defined bins of the outcome space, such as low, medium, and high values \citep{randahl2025bin}. 

For BCCP, non-conformity scores are calculated on a calibration data set within each user-defined bin. As the true bin is not known for new test points, for a new test point, the BCCP algorithm calculates its bin-specific non-conformity score within each bin.  A CP set is then constructed to include all bins for which the bin-specific conformal p-value of the test point exceeds $\alpha$.  For univariate outcomes, the final prediction set is formed by either the union of disjoint intervals corresponding to each bin in the CP set, or by contiguizing the intervals using the left- and right-most endpoints across the bins in the CP set. Both formations achieve \textit{at least} the desired coverage level, but the contiguized intervals may suffer over-coverage.  

BCCP is particularly useful in cases where the outcome space is skewed or imbalanced, as it allows for more flexible and adaptive CP sets that can better reflect the distribution of the outcome variable versus standard CP. It also allows the user to define the bins based on their knowledge of the data and the specific application, providing a more tailored approach to uncertainty quantification. Similar to MCP and CCP, BCCP requires carefully balancing the number and specification of bins with the quantity of data within each bin, to maximize statistical efficiency.

\subsection{Bootstrapped prediction intervals}

Outside the realm of CP, another widely used approach to constructing prediction intervals is based on bootstrapping. The bootstrap is a resampling technique that approximates the sampling distribution of any quantity of interest (e.g.~a point prediction) by repeatedly drawing samples with replacement from the observed data. This approach is particularly appealing when the distribution of prediction errors is unknown or complex, as it provides an empirical estimate of predictive uncertainty without requiring parametric assumptions \citep{beran1990calibrating, tian2022methods}.

Bootstrapped prediction intervals are expected to achieve nominal coverage under the assumption that the calibration set is representative of the data-generating process, and that the prediction errors are approximately independent and identically distributed. The algorithm proceeds by first computing the prediction errors on the calibration set and then repeatedly resampling these errors with replacement to form an empirical error distribution \citep{tian2022methods, hesterberg2015teachers}. For each new test observation, a large number of resampled errors are added to the model's point prediction, generating a simulated distribution of possible outcomes. The lower and upper bounds of the prediction interval are then obtained by taking $\hat{q}_{\alpha/2}$ and $\hat{q}_{1-\alpha/2}$. 

Several variants of the bootstrap exist, including the residual bootstrap, wild bootstrap, and parametric bootstrap.  Variants differ both in how the resampling is performed, and in the assumptions they require about the underlying model or error structure. 

\subsection{Parametric prediction intervals}
Parametric prediction intervals are constructed under the assumption that the prediction errors follow a specific probability distribution. This approach models uncertainty analytically rather than empirically, allowing the interval bounds to be derived directly from the assumed distribution and its estimated parameters. Common choices include the normal, log-normal, Poisson, or negative binomial distributions, depending on the nature of the outcome variable and the model's residuals.

For example, if the prediction errors are assumed to follow a centered normal distribution, their standard deviation can be estimated from a calibration set, and the prediction interval for a new observation is obtained by adding and subtracting the appropriate quantiles of the normal distribution around the point prediction. More generally, any distribution for which quantile functions are available can be used to construct prediction intervals, including user-specified or mixture distributions.

The main advantage of parametric prediction intervals lies in their computational efficiency and smooth, easily interpretable behavior. However, their validity depends critically on the appropriateness of the assumed error distribution: when this assumption is violated the resulting intervals may exhibit biased or unreliable coverage.

\section{The pintervals package and main functions}

The \pkg{pintervals} package provides a unified and model-agnostic interface for constructing prediction intervals using the methods described above. All functions in the package share a consistent syntax and workflow, allowing users to easily switch between different approaches and apply them to a wide range of predictive models. Each function takes as input a vector of predicted values for the test set, as well as predicted and true outcomes from calibration data\footnote{Calibration data are optional for parametric prediction intervals, as parameters can be supplied directly}, either as a two-column tibble or as separate vectors, and returns a tibble with the lower and upper bounds of the prediction intervals, along with the corresponding calibrated or uncalibrated predictions.  MCP- and CCP-based functions also return the group or cluster assignment for each observation.

For methods that rely on partitioning the data, such as MCP, CCP, or BCCP, an additional grouping variable must be specified. This can be provided either as a separate vector or as a column in the calibration tibble.

Prediction intervals are constructed using the \code{pinterval_*()} function family, which includes the following main functions:

\begin{itemize}
\item \code{pinterval_conformal()} for standard conformal prediction intervals
\item \code{pinterval_mondrian()} for Mondrian conformal prediction intervals
\item \code{pinterval_ccp()} for clustered conformal prediction intervals
\item \code{pinterval_bccp()} for bin-conditional conformal prediction intervals
\item \code{pinterval_bootstrap()} for bootstrapped prediction intervals
\item \code{pinterval_parametric()} for parametric prediction intervals
\end{itemize}

All functions follow the same general structure and support additional arguments to control key aspects of interval construction—such as the desired coverage level, the number of bootstrap replications, or the distributional parameters for parametric intervals. This unified design facilitates flexible experimentation and direct comparison across different interval estimation methods within a common analytical framework. The output of each function is a tibble containing the point predictions along with the lower and upper bounds of the prediction intervals and any additional relevant information (e.g., group or cluster assignments for MCP and CCP).

\subsection{Conformal prediction intervals}

The \code{pinterval_conformal()} function implements standard CP intervals. It takes the predicted values from the test set and the calibration data as input, and returns a tibble with the lower and upper bounds of the prediction intervals. The calibration data should be a two-column tibble of predicted (first column) and true (second column) values from the calibration set; alternatively the predicted values can be passed as a vector for the argument \code{calib} and the true values as a vector for the argument \code{calib_true}.

As additional arguments, the user can specify the desired coverage level and the non-conformity score function to be used. Absolute errors are used by default, with relative, zero-adjusted relative, heterogeneous, and raw errors as the alternatives. The user can also specify a custom real-valued function that takes the predicted and true values as input and returns a vector of non-conformity scores. This allows for flexibility in defining the non-conformity score, based on the specific application or data characteristics.

\subsubsection{Mondrian and clustered conformal prediction intervals}

The functions \code{pinterval_mondrian()} and \code{pinterval_ccp} implement MCP and CCP, which allows for partitioning the calibration set into disjoint groups based on a grouping variable.



The syntax for MCP and CCP builds on \code{pinterval_conformal()}, but requires an additional column in the calibration tibble for the grouping variable (third column). The grouping variable can also be passed as a vector for the argument \code{calib_group}.

For CCP, the user can also define the number of clusters to be used, or choose to optimize the number of clusters by minimizing the Cali{\'n}ski-Harabasz index of the proposed clustering structure \citep{calinski1974dendrite}. To avoid overfitting of the clusters, the calibration data can be further divided into one partition which is used to cluster the classes, and one partition used to calculate the resulting prediction intervals.\footnote{For more details on guidlines for how to divide the data, see \citet{ding2023class}.}

\subsubsection{Bin-conditional conformal prediction intervals}

The function \code{pinterval_bccp()} implements BCCP intervals, which allows for partitioning the calibration set into user-defined bins of the outcome space. The syntax is similar to \code{pinterval_mondrian()} and \code{pinterval_ccp()}, in that it uses an additional column, \code{bins}, in the calibration tibble to specify the bins, or a vector for the argument \code{calib_bins}. The user can also specify whether to contiguize the intervals using the argument \code{contiguize}, which is set to \code{FALSE} by default. If set to \code{TRUE}, the function will return contiguous intervals by taking the left- and right-most endpoints across the bins, otherwise it will contiguize adjacent bin-specific intervals. When \code{contiguize} is \code{FALSE} the function outputs a tibble where the prediction intervals with corresponding lower- and upper bounds are in a list-column called \code{intervals}, otherwsie the function outputs a standard tibble with point predictions and the contiguized lower- and upper bounds of the intervals. The user can also specify the desired coverage level and the non-conformity score function to be used, as in \code{pinterval_conformal()}.

\subsubsection{Distance weighted conformal prediction intervals}

All CP methods in \pkg{pintervals} also support distance-weighted CP (DWCP), which gives more weight to non-conformity scores from observations in the calibration set with values close to the new test observation. 

DWCP is enabled by setting \code{distance_weighted_cp = TRUE}. DWCP additionally requires calibration and test covariates, supplied through the arguments \code{distance_features_calib} and \code{distance_features_pred}. Distances are calculated as either the Mahalanobis distance (default) or the Euclidean distance between the rows of the calibration and test covariate data. Optionally, the distances can be normalized after calculation (preferred for Euclidean distances), either through min-max normalization or by dividing by the standard deviation of the distances (z-score normalization). When distance-weighted CP is used, the non-conformity scores from the calibration set are weighted according to their distance to the new test observation before computing the quantiles for the prediction interval.

In addition, the user can specify a weight function, which translates the distances into weights for the non-conformity scores. By default, the function uses the Gaussian kernel: $w = \exp(-d^2)$, where $d$ is the euclidean distance between the predicted value of the new test observation and the predicted values in the calibration set. Built in alternatives are the Cauchy kernel: $w = \frac{1}{1+d^2}$, the logistic kernel: $w = \frac{1}{1+\exp(d)}$, and the reciprocal linear kernel: $w = \frac{1}{1+d}$. The user can also specify a custom distance weighting function using the argument \code{distance_weighting_function}. This function should take a vector of distances as input and return a vector of non-negative weights, which will be used to weight the non-conformity scores. This allows for considerable flexibility in defining the distance weighting function based on the specific application or data characteristics.

\subsection{Bootstrapped prediction intervals}

The \code{pinterval_bootstrap()} function implements bootstrapped prediction intervals. It uses the prediction errors from the calibration set to mimic the distribution of errors via resampling. The function takes as input the predicted values for the test set and the calibration data, either as a two-column tibble (predicted values in the first column and true values in the second), or as separate vectors supplied to the arguments \code{calib} and \code{calib_truth}.


As additional arguments, the user can specify the desired coverage level (default is 0.9), the number of bootstrap samples (default is 1000), and the type of error to bootstrap from. The argument \code{error_type} can be set to \code{"raw"} to use raw, signed, errors, or \code{"absolute"} to use absolute errors with random signs. The user can also specify whether to use distance-weighted resampling by setting the argument \code{distance_weighted} to \code{TRUE}. This will give more weight to errors from observations in the calibration set with predicted values close to the new test observation, which can help improve the accuracy of the prediction intervals. The distance weighted bootstrap procedure follow the same principles as distance weighted CP described above, being controlled by the arguments \code{distance_features_calib}, \code{distance_features_pred}, \code{distance_type}, \code{normalize_distance}, and \code{weight_function}.

\subsection{Parametric prediction intervals}

The \code{pinterval_parametric()} function implements parametric prediction intervals based on a user-specified probability distribution. The user supplies the predicted values for the test set, a desired distribution (e.g.~normal, log-normal, Poisson, negative binomial), and either a calibration set with predicted and true values \textit{or} the parameters of the distribution.

Natively supported distributions include common continuous and count distributions such as the normal, log-normal, Poisson, and negative binomial distributions. The user specifies the distribution using the argument \code{dist}. For native distributions, the relevant parameters of the distribution (e.g.~standard deviation from the normal distribution and dispersion parameter from the negative binomial distribution) are estimated from the calibration set. Alternatively, the user can provide the parameters directly by passing a named list with the parameters through the argument \code{pars}. Unlike the other \code{pinterval_*()} functions, \code{pinterval_parametric()} does not require calibration data if the parameters of the distribution are provided directly. Similarly, for one-parameter distributions (e.g.~the Poisson and Chi-square distributions), no calibration data are needed as the single parameter can be estimated directly from the predicted values.

Beyond the natively supported distributions, the user can also use any arbitrary distribution by supplying a custom quantile function to the argument \code{dist}. This allows for considerable flexibility, including custom or mixture distributions using for instance the \pkg{mistr} package \citep{mistrpkg,mistrarticle}. When specifying a custom distribution, the user needs to provide the parameters of the distribution as a named list to the argument \code{pars}. Ideally, these parameters should be estimated on a calibration data set. The function will then use the specified quantile function to compute the lower and upper bounds of the prediction intervals.

Parametric intervals are simple, efficient and interpretable, but rely on correct specification of the underlying distribution. Users are encouraged to verify distributional assumptions using the calibration errors before applying this method.

\section{Using pintervals: An example with predicting county-level turnout}

To illustrate the practical use of the \pkg{pintervals} package, we demonstrate how it can be applied to generate prediction intervals for county-level voter turnout in the 2016 U.S. presidential election. This example provides a concrete, real-world application of the methods introduced above, using a typical regression problem workflow where predictions are obtained as point-predictions from a fitted model and prediction intervals are constructed using the methods available in \pkg{pintervals} to quantify the uncertainty around these predictions.

The data are bundled with the \pkg{pintervals} package and contains county-level information on voter turnout in the 2016 U.S. presidential election, along with various demographic and socioeconomic covariates that may influence turnout rates from the \citet{mitdata} dataset. The dataset includes variables such as population size, median income, education levels, as well as racial, age, and gender composition of the counties. The outcome variable of interest is the voter turnout rate, defined as the proportion of eligible voters who cast a ballot in the election. The dataset also includes a point prediction of turnout rates obtained from a random forest regression model using all available covariates, except geographic features such as state and latitude/longitude coordinates. The point predictions were generated using a leave-one-out cross-validation procedure to ensure that all the predictions are out-of-sample.\footnote{See the documentation of the dataset in \citet{pintervalspkg} for more details.}

The goal of the analysis is to construct prediction intervals around the point predictions of voter turnout rates using the different methods implemented in \pkg{pintervals} and evaluate their coverage and width. We will compare the performance of the different CP methods (standard, Mondrian, clustered, and bin-conditional), as well as bootstrapped prediction intervals and parametric prediction intervals assuming normal and beta distributions of errors.

We begin the example below by loading the necessary packages, the county turnout dataset, and splitting the data into calibration and test sets. We use a 50/50 random split of the data into calibration and test sets to illustrate the procedure. In practice, one would typically use a larger calibration set to ensure accurate estimation of the prediction intervals. Importantly, the point predictions used for calibration need to be generated from out-of-sample from the model to ensure valid coverage guarantees.\footnote{The point predictions in the dataset are generated using leave-one-out cross-validation to ensure they are out-of-sample.}

\begin{CodeChunk}
\begin{CodeInput}
R> library(pintervals)
R> library(dplyr)
R> library(tidyr)
R> data("county_turnout", package = "pintervals")
R> # Split data into calibration and test sets
R> set.seed(20260106) 
R> nobs <- nrow(county_turnout)
R> calib_indices <- sample(nobs, size = 0.5 * nobs)
R> calib_data <- county_turnout[calib_indices, ]
R> test_data <- county_turnout[-calib_indices, ]
\end{CodeInput}
\end{CodeChunk}

We can now proceed to construct prediction intervals using the different methods available in \pkg{pintervals}. For each method, we will use the point predictions from the random forest model and the true turnout rates from the calibration set to generate the intervals for the test set. We will then evaluate the empirical coverage and average width using the \code{interval_coverage()} function from the package.

\begin{CodeChunk}
\begin{CodeInput}
R> # Standard CP intervals
R> conformal_intervals <- pinterval_conformal(
 +   pred = test_data$predicted_turnout,
 +   calib = calib_data$predicted_turnout,
 +   calib_truth = calib_data$turnout,
 +   alpha = 0.1
 + )
R> # Evaluate coverage and width
R> conformal_coverage <- interval_coverage(
 +   truth = test_data$turnout,
 +   lower_bound = conformal_intervals$lower_bound,
 +   upper_bound = conformal_intervals$upper_bound)
R> conformal_coverage
\end{CodeInput}
\begin{CodeOutput}
[1] 0.9086229
\end{CodeOutput}
\end{CodeChunk}

We can repeat this procedure for the parametric and bootstrapped prediction intervals, adjusting the function calls as needed to accommodate the specific requirements of each method.

\begin{CodeChunk}
\begin{CodeInput}
R> # Parametric prediction intervals assuming normal distribution
R> norm_intervals <- pinterval_parametric(
 +   pred = test_data$predicted_turnout,
 +   calib = calib_data$predicted_turnout,
 +   calib_truth = calib_data$turnout,
 +   dist = "norm",
 +   alpha = 0.1
 + )
 +
R> # Evaluate coverage
R> norm_coverage <- interval_coverage(
 +   truth = test_data$turnout,
 +   lower_bound = norm_intervals$lower_bound,
 +   upper_bound = norm_intervals$upper_bound)
 +
R> # Parametric prediction intervals assuming logistic distribution
R> logis_intervals <- pinterval_parametric(
 +   pred = test_data$predicted_turnout,
 +   calib = calib_data$predicted_turnout,
 +   calib_truth = calib_data$turnout,
 +   dist = "logis",
 +   alpha = 0.1
 + )
 +
R> # Evaluate coverage
R> logis_coverage <- interval_coverage(
 +   truth = test_data$turnout,
 +   lower_bound = logis_intervals$lower_bound,
 +   upper_bound = logis_intervals$upper_bound)
 +
R> # Bootstrapped prediction intervals
R> bootstrap_intervals <- pinterval_bootstrap(
 +   pred = test_data$predicted_turnout,
 +   calib = calib_data$predicted_turnout,
 +   calib_truth = calib_data$turnout,
 +   alpha = 0.1,
 +   n_bootstrap = 1000
 + )
 +
R> # Evaluate coverage
R> bootstrap_coverage <- interval_coverage(
 +   truth = test_data$turnout,
 +   lower_bound = bootstrap_intervals$lower_bound,
 +   upper_bound = bootstrap_intervals$upper_bound)
 +
R> c(norm_coverage, logis_coverage, bootstrap_coverage)
\end{CodeInput}
\begin{CodeOutput}
[1]  0.9105534 0.9079794 0.9066924
\end{CodeOutput}
\end{CodeChunk}

Here we can see that all methods achieve coverage close to the nominal 90\% level. In the appendix, we present a more extensive simulation study comparing the performance of these methods across multiple calibration-test splits. These results show that across simulations, the SCP method achieve the most consistent coverage close to the nominal level, while the parametric methods slightly over-cover on average and the bootstrapped intervals slightly under-cover.

\subsection{Subgroup coverage with Mondrian and clustered CP}

Next, we demonstrate how to use Mondrian and clustered CP to improve coverage within subsets of the data. In this example, we will use the geographic group partition for the counties as the grouping variable for both methods. We will generate prediction intervals using Mondrian and clustered CP, as well as distance-weighted CP using geographic distances between counties. We will then evaluate the empirical coverage both in aggregate and within each of the four U.S. Census regions. For the CCP, we use the U.S. Census divisions as the grouping variable and optimize the number of clusters based on the Cali{\'n}ski-Harabasz index, setting the maximum number of clusters to 5.

\begin{CodeChunk}
\begin{CodeInput}
R> # MCP intervals
R> mondrian_intervals <- pinterval_mondrian(
 +   pred = test_data$predicted_turnout,
 +	pred_class = test_data$region,
 +   calib = calib_data$predicted_turnout,
 +   calib_truth = calib_data$turnout,
 +   calib_class = calib_data$region,
 +   alpha = 0.1
 + )
R> # Evaluate coverage
R> mondrian_coverage <- interval_coverage(
 +   truth = test_data$turnout,
 +   lower_bound = mondrian_intervals$lower_bound,
 +   upper_bound = mondrian_intervals$upper_bound)
 +
R> # Clustered CP intervals
R> clustered_conformal_intervals <- pinterval_ccp(
 +   pred = test_data$predicted_turnout,
 +	 pred_class = test_data$division,
 +   calib = calib_data$predicted_turnout,
 +   calib_truth = calib_data$turnout,
 +   calib_class = calib_data$division,
 +   alpha = 0.1,
 +   optimize_n_clusters = TRUE,
 +   max_n_clusters = 5,
 + )
R> # Evaluate coverage
R> clustered_conformal_coverage <- interval_coverage(
 +   truth = test_data$turnout,
 +   lower_bound = clustered_conformal_intervals$lower_bound,
 +   upper_bound = clustered_conformal_intervals$upper_bound)
 +
R> # Distance-weighted CP intervals using geographic distances
R> dw_cp_intervals <- pinterval_conformal(
 +   pred = test_data$predicted_turnout,
 +   calib = calib_data$predicted_turnout,
 +   calib_truth = calib_data$turnout,
 +   alpha = 0.1,
 +   distance_weighted_cp = TRUE,
 +   distance_features_calib = calib %>%
 +     select(latitude, longitude),
 +   distance_features_pred = test_data %>%
 +     select(latitude, longitude),
 +	 normalize_distance = "sd"
 + )
R> # Evaluate coverage
R> dw_cp_coverage <- interval_coverage(
 +   truth = test_data$turnout,
 +   lower_bound = dw_cp_intervals$lower_bound,
 +   upper_bound = dw_cp_intervals$upper_bound))
R> c(mondrian_coverage, clustered_conformal_coverage, dw_cp_coverage)
\end{CodeInput}
\begin{CodeOutput}
[1] 0.8931789 0.9021879 0.9054054
\end{CodeOutput}
\end{CodeChunk}

We again see that the methods achieve coverage close to the nominal 90\% level in aggregate. To evaluate the group-wise coverage within each U.S. Census region, we combine the results from all CP methods and compute the empirical coverage within each group.

\begin{CodeChunk}
\begin{CodeInput}
R> # Adding grouping variable and true turnout values
R> conformal_intervals <- conformal_intervals %>%
 +   mutate(region = test_data$region,
 +          method = "SCP",
 +          turnout = test_data$turnout)
R> mondrian_intervals <- mondrian_intervals %>%
 +   mutate(region = test_data$region,
 +					method = "MCP",
 +          turnout = test_data$turnout)
R> clustered_conformal_intervals <- clustered_conformal_intervals %>%
 +   mutate(region = test_data$region,
 +					method = "CCP",
 +          turnout = test_data$turnout)
R> dw_cp_intervals <- dw_cp_coverage %>%
 +   mutate(region = test_data$region,
 +					method = "DWCP",
 +          turnout = test_data$turnout)
R> bootstrap_intervals <- bootstrap_intervals %>%
 +   mutate(region = test_data$region,
 +					method = "Bootstrap",
 +          turnout = test_data$turnout)
R> norm_intervals <- norm_intervals %>%
 +   mutate(region = test_data$region,
 +					method = "Normal",
 +          turnout = test_data$turnout)
R> logis_intervals <- logis_intervals %>%
 +   mutate(region = test_data$region,
 +					method = "Logistic",
 +          turnout = test_data$turnout)
 +
R> # Combine all intervals for group-wise coverage evaluation
R> all_conformal_intervals <- bind_rows(
 +   conformal_intervals,
 +   mondrian_intervals,
 +   clustered_conformal_intervals,
 +   dw_cp_intervals,
 +   bootstrap_intervals,
 +   norm_intervals,
 +   logis_intervals
 + )
R> # Evaluate group-wise coverage
R> group_wise_coverage <- all_conformal_intervals %>%
 +   group_by(method, region) %>%
 +   summarise(
 +     coverage = interval_coverage(
 +       truth = turnout,
 +       lower_bound = lower_bound,
 +       upper_bound = upper_bound
 +     ),
 +     .groups = "drop"
 +   )
R> group_wise_coverage %>%
 +    pivot_wider(
 +     names_from = region,
 +     values_from = coverage
 +   )
\end{CodeInput}
\begin{CodeOutput}
 A tibble: 4 × 8
  region Bootstrap   CCP  DWCP Logistic   MCP Normal   SCP
  <chr>      <dbl> <dbl> <dbl>    <dbl> <dbl>  <dbl> <dbl>
1 Midwest    0.952 0.896 0.925    0.950 0.876  0.954 0.950
2 Northeast  0.929 0.902 0.902    0.946 0.902  0.946 0.946
3 South      0.884 0.904 0.890    0.884 0.899  0.886 0.885
4 West       0.862 0.913 0.913    0.867 0.913  0.867 0.867
\end{CodeOutput}
\end{CodeChunk}

The results show that the Mondrian, clustered, and distance-weighted CP algorithms achieve coverage that is more uniform and close to the nominal level across all U.S. Census regions compared to standard CP, bootstrapped, and parametric prediction intervals.

Analyzing the results further, we below compute the mean absolute error (MAE) of coverage across regions for each method to quantify the uniformity of coverage. The results of this analysis show that the Mondrian, clustered, and distance-weighted CP have MAE values around half of the alternative methods. The simulation study in the appendix further supports these findings, showing that the group-aware and distance-weighted CP methods consistently achieve coverage closer to the nominal level within the U.S. Census regions.

\begin{CodeChunk}
\begin{CodeInput}
R> group_wise_coverage %>%
 +	group_by(method) %>%
 +	summarize(mae_coverage = mean(abs(coverage - 0.9))) %>% 
 +    pivot_wider(
 +		names_from = method,
 +		values_from = mae_coverage
 +	)
\end{CodeInput}
\begin{CodeOutput}
 A tibble: 1 × 7
  Bootstrap     CCP   DWCP Logistic     MCP Normal    SCP
      <dbl>   <dbl>  <dbl>    <dbl>   <dbl>  <dbl>  <dbl>
1    0.0338 0.00579 0.0122   0.0365 0.00994 0.0368 0.0361

\end{CodeOutput}
\end{CodeChunk}

\subsection{Coverage within ranges of the target with bin-conditional CP}

Finally, we illustrate how BCCP can be used to improve coverage within different ranges of the outcome variable. In this example, we will create bins for turnout rates lower than 50\%, between 50\% and 60\%, between 60\% and 65\%, and above 65\%\footnote{These bins were chosen to roughly match the quartiles of turnout rates in the calibration set. In practice, the bins should be defined based on domain knowledge and the specific application at hand.} and generate prediction intervals using BCCP. We will then evaluate the empirical coverage both in aggregate and within each bin.

\begin{CodeChunk}
\begin{CodeInput}
R> # Create bins of turnout rates in the calibration set
R> calib_data <- calib_data %>%
 +   mutate(turnout_bin = case_when(
 +     turnout < 0.5 ~ 1,
 +     turnout < 0.6 ~ 2,
 +     turnout < 0.65 ~ 3,
 +     TRUE ~ 4
 +   ))
 +
R> bccp_contiguized_intervals <- pinterval_bccp(
 +   pred = test_data$predicted_turnout,
 +   calib = calib_data$predicted_turnout,
 +   calib_truth = calib_data$turnout,
 +   calib_bins = calib_data$turnout_bin,
 +   alpha = 0.1,
 +   contiguize = TRUE
 + )
 +
R> # Evaluate coverage
R> bccp_contiguized_coverage <- interval_coverage(
 +   truth = test_data$turnout,
 +   lower_bound = bccp_contiguized_intervals$lower_bound,
 +   upper_bound = bccp_contiguized_intervals$upper_bound)
 +
R> bccp_discontigous_intervals <- pinterval_bccp(
 +   pred = test_data$predicted_turnout,
 +   calib = calib_data$predicted_turnout,
 +   calib_truth = calib_data$turnout,
 +   calib_bins = calib_data$turnout_bin,
 +   alpha = 0.1,
 +   contiguize = FALSE
 + )
 +
R> # Evaluate coverage
R> bccp_discontigous_coverage <- interval_coverage(
 +   truth = test_data$turnout,
 +   intervals = bccp_discontigous_intervals$intervals
 + )
 +
R> c(bccp_contiguized_coverage, bccp_discontigous_coverage)
\end{CodeInput}
\begin{CodeOutput}
[1] 0.9026227 0.9018326
\end{CodeOutput}
\end{CodeChunk}

Both contiguized and non-contiguized BCCP intervals achieve coverage close to the nominal 90\% level in aggregate. To evaluate the bin-wise coverage within each turnout bin, we combine the results from both methods and compute the empirical coverage within each bin.

\begin{CodeChunk}
\begin{CodeInput}
R> # Adding bin variable to the intervals
R> test_data <- test_data %>%
 +   mutate(turnout_bin = case_when(
 +     turnout < 0.5 ~ 1,
 +     turnout < 0.6 ~ 2,
 +     turnout < 0.65 ~ 3,
 +     TRUE ~ 4
 +   ))
 +
R> bccp_contiguized_intervals <- bccp_contiguized_intervals %>%
 +   mutate(turnout_bin = test_data$turnout_bin,
 +          method = "BCCP(c)",
 +          turnout = test_data$turnout)
 +
R> bccp_discontigous_intervals <- bccp_discontigous_intervals %>%
 +   mutate(turnout_bin = test_data$turnout_bin,
 +          method = "BCCP(d)",
 +          turnout = test_data$turnout)
 +
R> conformal_intervals <- conformal_intervals %>%
 +   mutate(turnout_bin = test_data$turnout_bin)
 +
R> bootstrap_intervals <- bootstrap_intervals %>%
 +   mutate(turnout_bin = test_data$turnout_bin)
 +
R> norm_intervals <- norm_intervals %>%
 +   mutate(turnout_bin = test_data$turnout_bin)
 +
R> logis_intervals <- logis_intervals %>%
 +   mutate(turnout_bin = test_data$turnout_bin)
 +
R> # Combine all intervals for bin-wise coverage evaluation
R> all_intervals_bins <- bind_rows(
 +   bccp_contiguized_intervals,
 +   bccp_discontigous_intervals,
 +   conformal_intervals,
 +   bootstrap_intervals,
 +   norm_intervals,
 +   logis_intervals
 + )
 +
R> # Evaluate bin-wise coverage
R> bin_wise_coverage <- all_intervals_bins %>%
 +   group_by(method, turnout_bin) %>%
 +   summarise(
 +     coverage = interval_coverage(
 +       truth = turnout,
 +       lower_bound = lower_bound,
 +       upper_bound = upper_bound,
 +    	 intervals = intervals
 +     ),
 +     .groups = "drop"
 +   )
 +
R> bin_wise_coverage %>%
 +    pivot_wider(
 +     names_from = turnout_bin,
 +     values_from = coverage
 +   )
\end{CodeInput}
\begin{CodeOutput}

# A tibble: 6 × 5
  method      `1`   `2`   `3`   `4`
  <chr>     <dbl> <dbl> <dbl> <dbl>
1 BCCP(c)   0.909 0.927 0.921 0.904
2 BCCP(d)   0.909 0.915 0.918 0.904
3 Bootstrap 0.806 0.969 0.944 0.854
4 Logistic  0.814 0.969 0.941 0.856
5 Normal    0.814 0.971 0.944 0.861
6 SCP       0.814 0.969 0.941 0.859

\end{CodeOutput}

\end{CodeChunk}

In this case, we see that both contiguized and non-contiguized BCCP achieve coverage closer to the nominal level within each turnout bin compared to standard CP, bootstrapped, and parametric prediction intervals.

Analyzing the results further, we below compute the MAE of coverage across bins for each method to quantify their performance. This analysis show that the MAE of coverage across bins is lowest for the contiguized BCCP method, followed closely by the non-contiguized version. The simulation study in the appendix further supports these findings, showing that BCCP consistently achieves coverage closer to the nominal level within different ranges of the outcome variable.

\begin{CodeChunk}
\begin{CodeInput}
R> bin_wise_coverage %>%
 +	group_by(method) %>%
 +	summarize(mae_coverage = mean(abs(coverage - 0.9))) %>%
 +	pivot_wider(
 +		names_from = method,
 +		values_from = mae_coverage
 + )
\end{CodeInput}
\begin{CodeOutput}

# A tibble: 1 × 6
  `BCCP(c)` `BCCP(d)` Bootstrap Logistic Normal    SCP
      <dbl>     <dbl>     <dbl>    <dbl>  <dbl>  <dbl>
1    0.0155    0.0116    0.0631   0.0598 0.0598 0.0592

\end{CodeOutput}
\end{CodeChunk}

\section{Conclusion}

This paper has introduced the \pkg{pintervals} package for \proglang{R}, a unified and extensible framework for constructing prediction intervals using conformal, bootstrapped, and parametric methods. The package is designed to lower the barrier for applied researchers and practitioners to incorporate formal uncertainty quantification into their predictive analyses. By offering a consistent syntax and flexible interface, \pkg{pintervals} allows users to move seamlessly between different approaches to interval construction while maintaining transparent and reproducible workflows.

Through an empirical example predicting county-level voter turnout in the 2016 U.S. presidential election, we illustrated how the package can be applied to real-world data, and how alternative methods differ in coverage performance and interval width. The example further demonstrated how extensions such as Mondrian, clustered, bin-conditional, and distance-weighted CP can address heterogeneity and local calibration challenges across different subgroup dimensions of the data or prediction target. The results confirm that conformal methods achieve empirical coverage close to the nominal level not only overall but also within meaningful subpopulations, an important property in applied research where distributional assumptions may fail or subgroup balance is imperfect. These results are further validated in a simulation study presented in the appendix, showing the robustness and reliability of the CP methods implemented in \pkg{pintervals}.

Beyond the example presented here, \pkg{pintervals} can be readily integrated into a wide range of predictive modeling workflows, from classical regression to complex machine learning pipelines. The design emphasizes interpretability and extensibility, making it straightforward to evaluate new methods or tailor existing ones to domain-specific needs.

We hope that \pkg{pintervals} will provide an accessible, principled, and reproducible way for researchers and analysts to quantify predictive uncertainty. By combining simplicity of use with rigorous statistical foundations, the package aims to bridge the gap between methodological innovation and practical implementation, enabling more-trustworthy inference and decision-making in applied predictive modeling.

\bibliography{pintervals}

@inproceedings{williams2024decision,
  title={Decision theory via model-free generalized fiducial inference},
  author={Williams, Jonathan P and Liu, Yang},
  booktitle={Belief Functions: Theory and Applications},
  pages={131--139},
	volume={14909},
  year={2024},
  organization={Springer}
}

@article{williams2025,
  title={Model-free generalized fiducial inference},
  author={Williams, Jonathan P},
  journal={arXiv preprint arXiv:2307.12472},
  year={2023}
}

@article{balch2019,
  title={Satellite conjunction analysis and the false confidence theorem},
  author={Balch, Michael Scott and Martin, Ryan and Ferson, Scott},
  journal={Proceedings of the Royal Society A},
  volume={475},
  number={20180565},
  year={2019}
}

@article{martin2019,
title = {False confidence, non-additive beliefs, and valid statistical inference},
journal = {International Journal of Approximate Reasoning},
volume = {113},
pages = {39-73},
year = {2019},
author = {Ryan Martin}
}

@article{carmichael2018,
  title={An exposition of the false confidence theorem},
  author={Carmichael, Iain and Williams, Jonathan P},
  journal={Stat},
  volume={7},
  number={1},
  pages={e201},
  year={2018},
  publisher={Wiley Online Library}
}

@article{hegre2019views,
  title={ViEWS: A political violence early-warning system},
  author={Hegre, H{\aa}vard and Allansson, Marie and Basedau, Matthias and Colaresi, Michael and Croicu, Mihai and Fjelde, Hanne and Hoyles, Frederick and Hultman, Lisa and H{\"o}gbladh, Stina and Jansen, Remco and others},
  journal={Journal of peace research},
  volume={56},
  number={2},
  pages={155--174},
  year={2019},
  publisher={SAGE Publications Sage UK: London, England}
}

@article{shafer2008tutorial,
  title={A tutorial on conformal prediction.},
  author={Shafer, Glenn and Vovk, Vladimir},
  journal={Journal of Machine Learning Research},
  volume={9},
  number={3},
  year={2008}
}

@misc{mitdata,
title = {U.S. General Elections 2018 - Analysis Dataset},
author = {{MIT Election Data and Science Lab}},
year = {2018},
howpublished = {\url{https://github.com/MEDSL/2018-elections-unoffical/blob/master/election-context-2018.csv}},
note = {Accessed: 2025-11-03}
}

@article{hjort2025uncertainty,
  title={Uncertainty quantification in automated valuation models with spatially weighted conformal prediction},
  author={Hjort, Anders and Hermansen, Gudmund Horn and Pensar, Johan and Williams, Jonathan P},
  journal={International Journal of Data Science and Analytics},
  pages={1--18},
  year={2025},
  publisher={Springer}
}

@article{ding2023class,
  title={Class-conditional conformal prediction with many classes},
  author={Ding, Tiffany and Angelopoulos, Anastasios and Bates, Stephen and Jordan, Michael and Tibshirani, Ryan J},
  journal={Advances in neural information processing systems},
  volume={36},
  pages={64555--64576},
  year={2023}
}

@article{guan2023localized,
  title={Localized conformal prediction: A generalized inference framework for conformal prediction},
  author={Guan, Leying},
  journal={Biometrika},
  volume={110},
  number={1},
  pages={33--50},
  year={2023},
  publisher={Oxford University Press}
}

@article{calinski1974dendrite,
  title={A dendrite method for cluster analysis},
  author={Cali{\'n}ski, Tadeusz and Harabasz, Jerzy},
  journal={Communications in Statistics-theory and Methods},
  volume={3},
  number={1},
  pages={1--27},
  year={1974},
  publisher={Taylor \& Francis}
}

@article{hjort2024clustered,
  title={Clustered conformal prediction for the housing market},
  author={Hjort, Anders and Williams, Jonathan P and Pensar, Johan},
  journal={Proceedings of Machine Learning Research},
  volume={230},
  pages={1--21},
  year={2024}
}

@book{vovk2022algorithmic,
  title={Algorithmic Learning in a Random World},
  author={Vovk, Vladimir and Gammerman, Alexander and Shafer, Glenn},
  year={2022},
  publisher={Springer Nature}
}

@article{hesterberg2015teachers,
  title={What teachers should know about the bootstrap: Resampling in the undergraduate statistics curriculum},
  author={Hesterberg, Tim C},
  journal={The american statistician},
  volume={69},
  number={4},
  pages={371--386},
  year={2015},
  publisher={Taylor \& Francis}
}

@article{zhang2020random,
  title={Random forest prediction intervals},
  author={Zhang, Haozhe and Zimmerman, Joshua and Nettleton, Dan and Nordman, Daniel J},
  journal={The American Statistician},
  year={2020},
  publisher={Taylor \& Francis}
}

@article{beran1990calibrating,
  title={Calibrating prediction regions},
  author={Beran, Rudolf},
  journal={Journal of the American Statistical Association},
  volume={85},
  number={411},
  pages={715--723},
  year={1990},
  publisher={Taylor \& Francis}
}

@article{hegre20242023,
  title={The 2023/24 VIEWS Prediction challenge: Predicting the number of fatalities in armed conflict, with uncertainty},
  author={Hegre, H{\aa}vard and Vesco, Paola and Colaresi, Michael and Vestby, Jonas and Timlick, Alexa and Kazmi, Noorain Syed and Lindqvist-McGowan, Angelica and Becker, Friederike and Binetti, Marco and Bodentien, Tobias and others},
  journal={Journal of Peace Research},
  pages={00223433241300862},
  year={2024},
  publisher={SAGE Publications Sage UK: London, England}
}

@article{tian2022methods,
  title={Methods to compute prediction intervals: A review and new results},
  author={Tian, Qinglong and Nordman, Daniel J and Meeker, William Q},
  journal={Statistical Science},
  volume={37},
  number={4},
  pages={580--597},
  year={2022},
  publisher={Institute of Mathematical Statistics}
}

@article{chadefaux2017conflict,
  title={Conflict forecasting and its limits},
  author={Chadefaux, Thomas},
  journal={Data Science},
  volume={1},
  number={1-2},
  pages={7--17},
  year={2017},
  publisher={SAGE Publications Sage UK: London, England}
}

@article{brandt2014evaluating,
  title={Evaluating forecasts of political conflict dynamics},
  author={Brandt, Patrick T and Freeman, John R and Schrodt, Philip A},
  journal={International Journal of Forecasting},
  volume={30},
  number={4},
  pages={944--962},
  year={2014},
  publisher={Elsevier}
}

@article{petropoulos2022forecasting,
  title={Forecasting: theory and practice},
  author={Petropoulos, Fotios and Apiletti, Daniele and Assimakopoulos, Vassilios and Babai, Mohamed Zied and Barrow, Devon K and Taieb, Souhaib Ben and Bergmeir, Christoph and Bessa, Ricardo J and Bijak, Jakub and Boylan, John E and others},
  journal={International Journal of forecasting},
  volume={38},
  number={3},
  pages={705--871},
  year={2022},
  publisher={Elsevier}
}

@article{raftery2016use,
  title={Use and communication of probabilistic forecasts},
  author={Raftery, Adrian E},
  journal={Statistical Analysis and Data Mining: The ASA Data Science Journal},
  volume={9},
  number={6},
  pages={397--410},
  year={2016},
  publisher={Wiley Online Library}
}

@Manual{bootpkg,
    title = {boot: Bootstrap R (S-Plus) Functions},
    author = {{Angelo Canty} and {B. D. Ripley}},
    year = {2024},
    note = {R package version 1.3-31}
  }

@Manual{forecastpkg,
    title = {{forecast}: Forecasting functions for time series and linear models},
    author = {Rob Hyndman and George Athanasopoulos and Christoph Bergmeir and Gabriel Caceres and Leanne Chhay and Mitchell O'Hara-Wild and Fotios Petropoulos and Slava Razbash and Earo Wang and Farah Yasmeen},
    year = {2025},
    note = {R package version 8.24.0},
    url = {https://pkg.robjhyndman.com/forecast/}
}

@Article{mlr3,
    title = {{mlr3}: A modern object-oriented machine learning framework in {R}},
    author = {Michel Lang and Martin Binder and Jakob Richter and Patrick Schratz and Florian Pfisterer and Stefan Coors and Quay Au and Giuseppe Casalicchio and Lars Kotthoff and Bernd Bischl},
    journal = {Journal of Open Source Software},
    year = {2019},
    month = {dec},
    doi = {10.21105/joss.01903},
    url = {https://joss.theoj.org/papers/10.21105/joss.01903}
}

@Manual{mistrpkg,
    title = {mistr: Mixture and Composite Distributions},
    author = {Lukas Sablica and Kurt Hornik},
    year = {2023},
    note = {R package version 0.0.6},
    url = {https://CRAN.R-project.org/package=mistr},
    doi = {10.32614/CRAN.package.mistr},
  }

@Article{mistrarticle,
    title = {{mistr: A Computational Framework for Mixture and Composite Distributions}},
    author = {Lukas Sablica and Kurt Hornik},
    journal = {{The R Journal}},
    year = {2020},
    volume = {12},
    number = {1},
    pages = {283--299},
    doi = {10.32614/RJ-2020-003},
    url = {https://journal.r-project.org/archive/2020/RJ-2020-003/index.html},
  }

@Manual{lcppkg,
    title = {LCP: Localized conformal prediction.},
    author = {Leying Guan},
    year = {2021},
    note = {R package version 1.0, commit 64e5166864379a5e53d55ca22f5fe01986cde631},
    url = {https://github.com/LeyingGuan/LCP},
  }

@Manual{marginaleffectspkg,
    title = {marginaleffects: Predictions, Comparisons, Slopes, Marginal Means, and Hypothesis Tests
},
    author = {Vincent Arel-Bundock and Noah Greifer and Grant McDermott and Etienne Bacher and  Andrew Heiss},
    year = {2025},
    note = {R package version 0.31.0},
    url = {https://CRAN.R-project.org/package=marginaleffects},
    doi = {10.32614/CRAN.package.marginaleffects}
  }

@Manual{conformalforecastpkg,
    title = {conformalForecast: Conformal Prediction Methods for Multistep-Ahead Time Series
Forecasting},
    author = {Xiaoqian Wang and Rob Hyndman},
    year = {2025},
    note = {R package version 0.1.0},
    url = {https://CRAN.R-project.org/package=conformalForecast},
    doi = {10.32614/CRAN.package.conformalForecast},
  }

@Manual{conformalbayespkg,
    title = {conformalbayes: Jackknife(+) Predictive Intervals for Bayesian Models},
    author = {Cory McCartan},
    year = {2025},
    note = {R package version 0.1.4},
    url = {https://CRAN.R-project.org/package=conformalbayes},
    doi = {10.32614/CRAN.package.conformalbayes},
  }

@Manual{statspkg,
    title = {R: A Language and Environment for Statistical Computing},
    author = {{R Core Team}},
    organization = {R Foundation for Statistical Computing},
    address = {Vienna, Austria},
    year = {2025},
    url = {https://www.R-project.org/}
  }

@Manual{pintervalspkg,
    title = {pintervals: Model Agnostic Prediction Intervals},
    author = {David Randahl and Jonathan P. Williams},
    year = {2025},
    note = {R package version 0.9.2},
  }

@Manual{tidymodelspkg,
    title = {Tidymodels: a collection of packages for modeling and machine learning using tidyverse principles.},
    author = {Max Kuhn and Hadley Wickham},
    url = {https://www.tidymodels.org},
    year = {2020}
  }

@article{hegre2021views2020,
  title={ViEWS2020: revising and evaluating the ViEWS political violence early-warning system},
  author={Hegre, H{\aa}vard and Bell, Curtis and Colaresi, Michael and Croicu, Mihai and Hoyles, Frederick and Jansen, Remco and Leis, Maxine Ria and Lindqvist-McGowan, Angelica and Randahl, David and R{\o}d, Espen Geelmuyden and others},
  journal={Journal of peace research},
  volume={58},
  number={3},
  pages={599--611},
  year={2021},
  publisher={Sage Publications Sage UK: London, England}
}

@book{arel2025model,
  title={Model to Meaning: How to Interpret Statistical Models with R and Python},
  author={Arel-Bundock, Vincent},
  year={2025},
  publisher={CRC Press}
}

@article{randahl2025bin, title={Bin-Conditional Conformal Prediction of Fatalities from Armed Conflict}, DOI={10.1017/pan.2025.10010}, journal={Political Analysis}, author={Randahl, David and Williams, Jonathan P. and Hegre, Håvard}, year={2025}, pages={1–13}}

@article{randahl2025forecasting,
title = {Forecasting electoral violence},
journal = {International Journal of Forecasting},
year = {2025},
issn = {0169-2070},
doi = {https://doi.org/10.1016/j.ijforecast.2025.09.003},
url = {https://www.sciencedirect.com/science/article/pii/S0169207025000871},
author = {David Randahl and Maxine Leis and Tim Gåsste and Hanne Fjelde and Håvard Hegre and Staffan I. Lindberg and Steven Wilson}
}

@article{lyu2021forecasting,
  title={Forecasting US economic growth in downturns using cross-country data},
  author={Lyu, Yifei and Nie, Jun and Yang, Shu-Kuei X},
  journal={Economics letters},
  volume={198},
  pages={109668},
  year={2021},
  publisher={Elsevier}
}

@article{nanlohy2017policy,
  title={The policy value of quantitative atrocity forecasting models},
  author={Nanlohy, Sascha and Butcher, Charles and Goldsmith, Benjamin E},
  journal={The RUSI Journal},
  volume={162},
  number={2},
  pages={24--32},
  year={2017},
  publisher={Taylor \& Francis}
}

@article{hegre2017introduction,
  title={Introduction: Forecasting in peace research},
  author={Hegre, H{\aa}vard and Metternich, Nils W and Nyg{\aa}rd, H{\aa}vard Mokleiv and Wucherpfennig, Julian},
  journal={Journal of Peace Research},
  volume={54},
  number={2},
  pages={113--124},
  year={2017},
  publisher={Sage Publications Sage UK: London, England}
}

@article{Barber2023,
author = {Foygel Barber, Rina and  Cand{\`e}s, Emmanuel and Ramdas, Aaditya and Tibshirani, Ryan J.},
title = {{Conformal prediction beyond exchangeability}},
volume = {51},
journal = {The Annals of Statistics},
number = {2},
publisher = {Institute of Mathematical Statistics},
pages = {816 -- 845},
year = {2023},
}

@inproceedings{Tibshirani2019,
 author = {Tibshirani, Ryan J and Foygel Barber, Rina and Candes, Emmanuel and Ramdas, Aaditya},
 booktitle = {Advances in Neural Information Processing Systems},
 title = {Conformal Prediction Under Covariate Shift},
 volume = {32},
 year = {2019}
}

@article{MartinMaoReich2023,
author = {Huiying Mao and Ryan Martin and Brian J. Reich},
title = {Valid Model-Free Spatial Prediction},
journal = {Journal of the American Statistical Association},
volume = {119},
number = {546},
pages = {904--914},
year = {2023},
}

@article{barber2025unifying,
  title={Unifying different theories of conformal prediction},
  author={Barber, Rina Foygel and Tibshirani, Ryan J},
  journal={arXiv preprint arXiv:2504.02292},
  year={2025}
}

@article{LeiWasserman2013,
    author = {Lei, Jing and Wasserman, Larry},
    title = {Distribution-free Prediction Bands for Non-parametric Regression},
    journal = {Journal of the Royal Statistical Society Series B: Statistical Methodology},
    volume = {76},
    number = {1},
    pages = {71-96},
    year = {2013},
    issn = {1369-7412},
    doi = {10.1111/rssb.12021},
    url = {https://doi.org/10.1111/rssb.12021},
    eprint = {https://academic.oup.com/jrsssb/article-pdf/76/1/71/49514328/jrsssb_76_1_71.pdf},
}

\newpage

\section*{Appendix: Simulation-based evaluation of prediction interval methods}

This appendix presents a simulation study designed to systematically evaluate the empirical coverage properties of the prediction interval methods implemented in the \pkg{pintervals} package. While the main text illustrates these methods using a single calibration-test split, the simulation framework assesses their performance across repeated random splits of the data. The focus is on three complementary dimensions of validity: aggregate coverage, subgroup-wise coverage, and coverage within ranges of the outcome variable. In total, we run 1000 simulation iterations for each setting, generating new calibration and test sets in each iteration.

Across all simulation settings, the nominal coverage level is fixed at $1 - \alpha = 0.9$. For each simulation run, prediction intervals are constructed using the same procedures described in the main text, and performance is summarized using (i) empirical coverage, taken as the proportion of test observations whose true outcome lies within the corresponding prediction interval, averaged across simulation runs, and (ii) the mean absolute error (MAE) of coverage relative to the nominal level. The MAE metric is defined as

$$\text{MAE} = \frac{1}{G} \sum_{g=1}^{G} \left| \hat{c}_g - (1 - \alpha) \right|$$

where $\hat{c}_g$ denotes empirical coverage in group g, and G is the number of groups or bins under consideration. Lower MAE values indicate more uniform and locally valid coverage. The MAE is computed for each simulation iteration and then averaged across all runs to summarize performance.

\subsection*{Aggregate coverage simulations}

The first set of simulations evaluates aggregate coverage, averaging across all observations in the test set without conditioning on subgroups or outcome ranges. For each simulation iteration, the data are randomly split into calibration and test sets, and prediction intervals are constructed using standard conformal prediction, parametric prediction intervals, and bootstrap-based intervals. Here, we compare the standard CP method against parametric intervals assuming normal and logistic distributions, as well as bootstrapped intervals.

The results show that all four methods achieve nominal coverage very close to the nominal level of 0.9, with MAE values below 0.01. Both of the parametric methods have slight over-coverage, while the bootstrap method has a minimal level of under-coverage.\footnote{These patterns were confirmed by rerunning the analysis using different random seeds, with the bootstrap method achieving coverage rates of 89.7--89.9\%, and the logistic method achieving coverage rates of 90.3--90.5\%, and the normal method slightly higher.} These findings confirm the robustness of CP methods in achieving valid coverage without relying on distributional assumptions.

\begin{table}[ht]
\centering
\begin{tabular}{lrr}
  \hline
Method & Mean Coverage & MAE Coverage \\ 
  \hline
 Logistic & 0.903 & 0.008 \\ 
 SCP & 0.899 & 0.009 \\ 
 Bootstrap & 0.898 & 0.009 \\ 
 Normal & 0.907 & 0.009 \\ 
   \hline
\end{tabular}
\end{table}

\subsection*{Subgroup-wise coverage simulations}
The second set of simulations evaluates coverage within predefined subgroups of the data. These tests are motivated by the concern that methods with correct aggregate coverage may still exhibit systematic under- or over-coverage within substantively meaningful groups.

In each simulation run, prediction intervals are constructed using standard conformal prediction (SCP), Mondrian conformal prediction (MCP), clustered conformal prediction (CCP), and distance-weighted conformal prediction (DWCP), as well as bootstrapped and parametric (logistic and normal) intervals. 

Group membership is defined using geographic classifications, corresponding to U.S. Census regions and divisions, as described in the main text. Coverage is computed separately within each group, while MAE of coverage is calculated across groups to quantify uniformity.

The results show that while all methods achieve empirical coverage close to the nominal level in aggregate, the group-aware CP methods (MCP, CCP, DWCP) have group-wise coverage much closer to the nominal level compared to standard CP, bootstrapped, and parametric intervals. Specifically, MCP, CCP, and DWCP. In addition, these methods exhibit substantially lower MAE of coverage across regions compared to standard CP, bootstrapped, and parametric intervals. This indicates that the group-aware methods provide more uniform and reliable coverage within subpopulations, addressing potential heterogeneity in the data.

\begin{table}[ht]
\centering
\begin{tabular}{lrrrr:rr}
  & \multicolumn{4}{c}{Empirical Coverage by Region} & \multicolumn{2}{c}{Overall Metrics} \\
  \cline{2-5} \cline{6-7} \\
Method & Midwest & Northeast & South & West & Mean Coverage & MAE Coverage \\ 
  \hline
MCP & 0.898 & 0.890 & 0.898 & 0.894 & 0.897 & 0.022 \\ 
CCP & 0.904 & 0.884 & 0.894 & 0.880 & 0.895 & 0.024 \\ 
DWCP & 0.942 & 0.901 & 0.877 & 0.883 & 0.901 & 0.028 \\ 
Bootstrap & 0.948 & 0.927 & 0.873 & 0.844 & 0.898 & 0.040 \\ 
Normal & 0.956 & 0.937 & 0.882 & 0.851 & 0.907 & 0.040 \\ 
SCP & 0.949 & 0.929 & 0.873 & 0.847 & 0.899 & 0.040 \\ 
Logistic & 0.952 & 0.934 & 0.877 & 0.849 & 0.903 & 0.040 \\ 
   \hline
\end{tabular}
\end{table}

\subsection*{Coverage within ranges of the target simulations}

The third set of simulations focuses on coverage within different ranges of the outcome variable. This scenario is particularly relevant when prediction errors vary systematically across the outcome space, such as in the tails of the distribution.

In each simulation iteration, the calibration data are partitioned into bins based on the observed outcome, using thresholds chosen to approximately balance the number of observations across four bins. Prediction intervals are then constructed using bin-conditional conformal prediction (BCCP), with both contiguized and non-contiguized variants, alongside standard conformal, parametric, and bootstrap-based intervals.

Coverage is computed separately within each outcome bin. As in the grouped simulations, performance is summarized using mean aggregate coverage and the MAE of bin-wise coverage relative to the nominal level.

The results indicate that both contiguized and non-contiguized BCCP achieve empirical coverage closer to the nominal level within each outcome bin compared to standard CP, bootstrapped, and parametric prediction intervals. The MAE of coverage across bins is lowest for the contiguized BCCP method, followed closely by the non-contiguized version. These findings highlight the effectiveness of BCCP in addressing heterogeneity in prediction errors across the outcome space, leading to more reliable uncertainty quantification. As expected, the aggregate coverage for the contiguized BCCP is slightly higher than nominal due to the effect of contiguization.

\begin{table}[ht]
\centering
\begin{tabular}{lrrrr:rr}
& \multicolumn{4}{c}{Empirical Coverage by Outcome Bin} & \multicolumn{2}{c}{Overall Metrics} \\
  \cline{2-5} \cline{6-7} \\
Method & 1 & 2 & 3 & 4 & Mean Coverage & MAE Coverage \\ 
  \hline
BCCP(d) & 0.895 & 0.898 & 0.899 & 0.894 & 0.897 & 0.019 \\ 
BCCP(c) & 0.895 & 0.909 & 0.912 & 0.894 & 0.906 & 0.020 \\ 
Normal & 0.804 & 0.960 & 0.937 & 0.779 & 0.907 & 0.079 \\ 
Logistic & 0.796 & 0.957 & 0.934 & 0.773 & 0.903 & 0.081 \\ 
Bootstrap & 0.796 & 0.954 & 0.929 & 0.760 & 0.898 & 0.082 \\ 
SCP & 0.788 & 0.955 & 0.932 & 0.768 & 0.899 & 0.083 \\ 
   \hline
\end{tabular}
\end{table}

\end{document}